\documentstyle[12pt]{article}
\begin{document}
\hsize = 6.0 in
\hoffset= -0.25 in
\voffset=-.5 in
\baselineskip=15pt
\newcommand{\eqletter}{ \hfill (\theequation\alph{letter})}
\newcommand{\gm}{{(\Box+e^2\rho^2)}}
\newcommand{\eql}{\nonumber &\eqletter \cr
                  \addtocounter{letter}{1}}
\newcommand{\be}{\begin{equation}}
\newcommand{\ee}{\end{equation}}
\newcommand{\bea}{\begin{eqnarray}}
\newcommand{\eea}{\end{eqnarray}}
\newcommand{\beal}{\setcounter{letter}{1} \begin{eqnarray}}
\newcommand{\eeal}{\addtocounter{equation}{1} \end{eqnarray}}
\newcommand{\none}{\nonumber \\}
\newcommand{\vbar}{\overline{v}}
\newcommand{\xbar}{\overline{x}}
\newcommand{\req}[1]{Eq.(\ref{#1})}
\newcommand{\reqs}[1]{Eqs.(\ref{#1})}
\newcommand{\larrow}{\,\,\,\,\hbox to 30pt{\rightarrowfill}
\,\,\,\,}
\newcommand{\slarrow}{\,\,\,\hbox to 20pt{\rightarrowfill}
\,\,\,}
\newcommand{\half}{{1\over2}}
\newcommand{\Fstar}{\tilde{F}}
\begin{center}
{\bf
Massless Scalar QED}\\
{\bf with Non-Minimal Chern-Simons Coupling}
\\[20pt]
{\sl by}\\[20pt]
{M.E. Carrington and G. Kunstatter}\\[5pt]
{\sl
 Dept. of Physics and Winnipeg Institute of
 Theoretical Physics\\
University of Winnipeg, Winnipeg, Manitoba\\
Canada R3B 2E9
}
\end{center}
\vspace*{1.5cm}
\par
\noindent
{\bf ABSTRACT:} 2+1 dimensional massless scalar QED with
$(\phi^*\phi)^3$
scalar self-coupling is modified by the addition of a non-
minimal Chern-Simons term that couples the dual of the
electromagnetic field strength to the covariant current
of the complex scalar field. The theory is shown to be fully one-
loop renormalizable.  The one loop effective
potential for the scalar field gives rise to spontaneous symmetry
breaking which induces masses for  both the scalar and vector
fields.
At high temperature there is a symmetry restoring phase transition.
\\[60pt]
WIN-93-10
\\
{\it December, 1993}
\clearpage
\section{Introduction}
Lower dimensional field theories with Chern-Simons terms are of
interest for a variety of reasons.
They possess interesting theoretical properties\cite{cs} and have
been
proposed to describe the physics of planar systems relevant to the
fractional
Hall effect \cite{hall} and high temperature superconductivity
\cite{high T}.
Any theory that can describe superconductivity must contain a phase
transition or, in the language of quantum field theory, such a
theory must
exhibit
spontaneous symmetry breaking.  Previous work has attempted to
study the
superconducting phase transition by considering scalar quantum
electrodynamics
(QED) minimally coupled to a Chern-Simons (CS) term (see Burgess
{\it et al} \cite{burgess} and references therein).  This model is
a natural
place to start since scalar QED in $3+1$ dimensions exhibits
spontaneous
symmetry breaking and, in $2+1$ dimensions, the coupling to the CS
term gives
fractional statistics to the scalar fields.
The Lagrangian has the form:
\be
{\cal L}= \half(D_\mu \phi)^* (D^\mu \phi) - \half
\mu^2(\phi^*\phi)
-{\tau \over 4!} (\phi^*\phi)^2 - {\lambda\over 6!}
(\phi^*\phi)^3-{1\over4} F_{\mu\nu}
F^{\mu\nu} + {m_{CS} \over 4}A^\mu \Fstar_\mu
\label{eq: oldlagrangian}
\ee
$\phi=(\chi+ i\eta)$ is a complex-valued
scalar field,
$\{{\mu, \nu....= 0, 1,2}\}$ are Lorentz indices and  the covariant
derivative is:
\be
D_\mu\phi = \partial_\mu\phi +ie A_\mu\phi\,\, .
\ee
$\Fstar^\mu[A] = \epsilon^{\mu\nu\alpha} F_{\nu\alpha}$ is
the dual of the electromagnetic field strength $F_{\mu\nu} =
\partial_\mu
A_\nu -\partial_\nu A_\nu$ with
$\epsilon^{\mu\nu\alpha}$ the totally antisymmetric tensor density,
$\epsilon^{012}=1$. A detailed analysis of the dynamics of the
phase transition in the above model\cite{burgess} has shown that
the CS term has essentially no effect other than that of shifting
the zero-point energy.  If the phase transition is to be
interpreted as a superconducting phase transition it seems
surprising that its physical properties are
independent of the term that is responsible for generating
fractional statistics.
\par
In the present paper
we will examine a model in which the scalar field is non-minimally
coupled to a manifestly gauge invariant CS term. In
particular, we will replace the standard CS term in the Lagrangian
\req{eq: oldlagrangian} by a term of the form,
\be
{\cal L}_{CS} =
- {i \over4} \gamma [|\phi|] \Fstar^\mu[A] J_\mu[A,\phi]
\label{eq: CSlagrangian}
\ee
The covariant electromagnetic current associated with the scalar
field in our model is,
\be
J_\mu[\phi,A]= \frac{1}{2}\left[\phi^* (D_\mu\phi) -
(D_\mu\phi)^* \phi\right]
\label{eq: current}
\ee
and
$\gamma[|\phi|]$ is an arbitrary function of the magnitude of the
scalar field.
This term is the only gauge invariant expression that
can be added to the Lagrangian without giving up one-loop
renormalizability.  We will show below that the standard CS term is, in fact,
contained in this more general expression.
As in standard
Abelian CS theory, ${\cal L}_{CS}$  is odd
under parity reversal, and is topological, in the sense that it
does not depend on the space time metric. In the present case, it
is also manifestly gauge
invariant. Its connection with the standard CS term is
most easily seen by going to the parametrization:
\be
\phi=|\phi|
e^{i\alpha}.
\label{eq: magnitude phase}
\ee
In this
case,
\be
{\cal L}_{CS}=\frac{1}{4} e\gamma[|\phi|]\phi^*\phi
(A_{\mu}+\frac{1}{e}\partial_\mu \alpha)
\Fstar^\mu \,\, .
\ee
If one chooses  $\gamma= m/(e\phi^*\phi)$,  then  ${\cal L}_{CS}$
is
a manifestly
gauge invariant form of the Abelian CS term with
topological mass $m$.
The second term, which is necessary for gauge invariance, is a total
divergence and can be dropped, leaving the standard CS Lagrangian.
\par
For $\gamma=k/|\phi|$, the theory is equivalent to one considered
by
Paul and
Khare\cite{pk} who started with a kinetic term for the scalar field
written
in
terms of a non-minimal covariant derivative:
\be
{\cal D}_\mu\phi =D_\mu \phi +ig\Fstar_\mu {\phi\over |\phi|}
\ee
The Lagrangian they used was:
\bea
\tilde{\cal L}&=& \half({\cal D}^\mu \phi)^* ({\cal D}_\mu \phi)
   -V(\phi) -\frac{1}{4} (1+g^2) F^{\mu\nu} F_{\mu\nu}\none
&=& \half(D_\mu \phi)^* (D_\mu \phi) - V(\phi) -{1\over4}
F_{\mu\nu}
F^{\mu\nu}  -\frac{ig}{|\phi|} \Fstar^\mu J_\mu
\label{eq: khare's lagrangian}
\eea
which is clearly equivalent to the one above with $\gamma=
4g/|\phi|$.
\par
In this paper we will examine in detail the quantum effects in
this theory at the one loop level. One of the more remarkable
results is that  renormalizability of the effective potential
puts severe restrictions on the form of $\gamma$. In fact, if we
restrict $\gamma$ to be a polynomial function of the magnitude of
the scalar field, then there are only two distinct possibilities:
One possibility is
\be
\gamma= \gamma_0 + m_{CS} / (e\phi^*\phi)
\label{eq: our gamma}
\ee
In this case, the CS part
of the Lagrangian \req{eq: CSlagrangian} has the simple
form:
\be
{\cal L}_{CS} = {1\over 4}m_{CS}A^\mu\Fstar_\mu
+ {1\over 4}\gamma_0 J^\mu\Fstar_\mu
\label{eq: simpleCS1}
\ee
Note that the first term in \req{eq: simpleCS1} is just the usual CS term.
Moreover, as will be shown in section 3, \req{eq: CSlagrangian} is the most
general expression, that includes the standard CS term, that can be added
to the Lagrangian without destroying one-loop renormalizability.
The second possibility is essentially the
Lagrangian of Paul and Khare\cite{pk}.  However, this choice is
somewhat unnatural in the sense that it introduces an explicit
dependence on  $|\phi|$ into a Lagrangian with
other terms that depend only on $\phi^*\phi$.
\par
In the following, we will therefore restrict consideration to
$\gamma$ of the form \req{eq: our gamma} given above.
Since our purpose is to consider a theory in which all
masses are radiatively
induced  we will also set the bare masses $\mu$ and $m_{CS}$ to
zero in
\req{eq: oldlagrangian} and \req{eq: simpleCS1}.  We will show that a finite
value for $m_{CS}$ would not effect the renormalizability of the theory, or the
effective potential except for a shift of the zero point energy.
The paper is organized as follows: In
section 2 we give the equations of motion, the conserved current, and the
Feynman rules for this theory.  In section 3
we show that the symmetric vacuum
is unstable against radiative corrections. The one loop effective
potential for the scalar field has a symmetry breaking
minimum which induces masses for both the scalar and vector modes
in the theory. As in the standard Coleman-Weinberg mechanism
\cite{cw}, the
resulting mass ratios can be expressed in terms of the only
dimensionless coupling constants in the theory, namely $\lambda$
and $\delta\equiv \gamma_0 e$.
In section 4  we derive the necessary one-loop counter terms and
show that they  are all of the form of terms found in
\req{eq: oldlagrangian}.
In section 5 we derive the finite temperature contribution to the
one loop
effective potential and show that there is a symmetry restoring
phase
transition at a high temperature. Finally, in section 6 we close with
a discussion of the results and prospects for future work.

\section{Feynman Rules}
\par
We start with the  bare Lagrangian,
\be
{\cal L}= \half(D_\mu \phi)^* (D^\mu \phi)  - {\lambda\over 6!}
(\phi^*\phi)^3-{1\over4} F_{\mu\nu}
F^{\mu\nu} - {i \over4} \gamma_0  \Fstar^\mu J_\mu
\label{eq: newlagrangian}
\ee
The conserved current associated with the gauge invariance is
\be {\cal J}_\mu = J_{\mu}+ \frac{i}{4}\gamma_0 \phi^*\phi
\tilde{F}_\mu
\ee
The equations of motion for the gauge boson field and the scalar field are,
\be
\partial^\tau F_{\tau\mu} = ie{\cal J}_{\mu} +\frac{i}{2}\gamma_0
\epsilon
_{\mu\tau\lambda}\partial^\tau  J^\lambda
\ee
\be
D^{\mu}D_{\mu}\phi + \frac{\lambda}{5!}(\phi^*\phi)^2\phi +
i\frac{\gamma_0}{2}\Fstar_{\mu}D^\mu \phi =0
\ee

Since we are
interested in generating masses for the fields by radiative
corrections, we have set the bares masses $\mu$ and $m_{CS}$ to zero.  Mass
counter-terms will appear, however, because there
is no symmetry of the theory that ensures that the bare parameters will vanish
in
the limit that the renormalized parameters are zero.  We  have verified that
none of our results would change if we had
used finite bare masses, other than the form of the mass counter-terms which
are necessary to give zero renormalized masses at $v=0$, where $v$ is the
vacuum
expectation value of the real part of the scalar field.  Thus, the values of
the bare masses have no physical significance.  We will obtain a renormalized
theory with radiatively induced spontaneous symmetry breaking by setting the
physical renormalized masses to zero at $v=0$, and using this equation as a
renormalization condition.  For simplicity, we have also chosen to
set the bare quartic coupling to zero.  We will derive the quartic counter-term
that is necessary to give a renormalized quartic
coupling of zero at $v=0$.

To quantize the theory we must add a gauge fixing term to the
Lagrangian in
\req{eq: newlagrangian}.  We use the $R_{\xi}$
gauge where power counting arguments are possible.
The gauge fixing term is,
\be
{\cal L}_{GF} = -\frac{1}{2\xi}(\partial_\mu A^\mu - \xi e v
\eta)^2
\label{eq: gflagrangian}
\ee

To study spontaneous symmetry breaking, we shift the real part of
the scalar
field: $\chi\rightarrow \chi + v$.
The Lagrangian becomes:
\bea
{\cal L} &=& {1\over 2} (\partial_\mu \chi)^2 + {1\over 2}
(\partial_\mu \eta)^2 -{1\over 2}
\xi e^2 v^2 \eta^2 -{1\over 4} F_{\mu\nu}F^{\mu\nu} -{1\over
{2\xi}}
(\partial_\mu
A^\mu)^2
\none
&+& {1\over 2} e^2v^2 g_{\mu\nu}A^\mu A^\nu  +
{1\over 2} \delta \epsilon^{\mu\nu\alpha}v^2
A_{\mu}\partial_{\nu}A_{\alpha}
+eA^{\mu}(\chi\partial_\mu \eta - \eta \partial_\mu \chi)
\none
&+&{1\over 2} e^2 g_{\mu\nu}A^\mu A^\nu (\eta^2+\chi^2)
+\frac{\gamma_0}{2} (\chi\partial_\mu \eta - \eta \partial_\mu \chi)
    \epsilon^{\mu\alpha\nu}\partial_{\alpha} A_\nu
+ \frac{\delta}{2} \epsilon^{\mu\alpha\nu}(\chi^2+\eta^2)A_\mu
\partial_{\alpha}
    A_{\nu}
\none
&+& \delta v \chi \epsilon^{\mu\alpha\nu}A_\mu\partial_\alpha A_{\nu}
+ e^2v\chi g_{\mu\nu}A^{\mu}A^\nu
\none
&-&\frac{\lambda}{6!}[\chi^6 + \eta^6 + 3\chi^4\eta^2 +
3\chi^2\eta^4
    + v(6\chi^5 + 12 \chi^3\eta^2 + 6\chi\eta^4)
    +v^2(15\chi^4 +3\eta^4 + 18\eta^2\chi^2)
\none
&+& v^3(20\chi^3 + 12\chi\eta^2)
    + v^4(15\chi^2 + 3\eta^2) ]
\eea
We obtain the propagators from the terms in the Lagrangian that are
quadratic
in the fields by inverting the appropriate operators.
We use the Landau value for the gauge fixing parameter $\xi = 0$
so that there
are no contributions from ghost degrees of freedom. This procedure
gives,
\bea
iS_\chi (P) &=& \frac{i}{P^2 -m_\chi ^2+i\epsilon}
\none
iS_\eta(P) &=& \frac{i}{P^2 -m_\eta ^2 +i\epsilon}
\eea
\be
-iD_{\mu\nu}(K) =
-i\frac{(K^2-e^2v^2)(g_{\mu\nu}K^2-K_{\mu}K_{\nu})}
{K^2(K^2-m_1^2)(K^2-m_2^2)}
- \frac{\delta v^2 K^{\alpha}\epsilon_{\mu\alpha\nu}}
{(K^2-m_1^2)(K^2-m_2^2)}\none
\ee
where $m_\chi ^2 = \frac{\lambda}{4!}v^4 $,
$m_\eta ^2 = \frac{\lambda}{5!}v^4 $ and,
\be
m^2_{1,2} = e^2v^2\left[ 1+\half\gamma_0^2 v^2 \pm\half\gamma_0 v
\sqrt{4+\gamma_0^2
v^2}\right]
\label{eq: m12}
\ee
When the vev of the
scalar field is
non-zero, the particle spectrum contains four massive modes.  All
the masses
are real as long as $\gamma_0$ is real.
 In the absence of the CS term
($\gamma_0=0$), the gauge boson masses  become degenerate:
$m_{1,2}=ev$.
Note that the Feynman rules would in general be more complicated if we
had used arbitrary $\gamma=\gamma[v]$ instead of the constant value
$\gamma=\gamma_0$.  However, the photon propagator is valid
for arbitrary
$\gamma=\gamma[v]$, and one can immediately obtain masses for standard CS
theory\cite{pr}  by simply
substituting $\gamma_0= m_{CS}/ev^2$ into \req{eq: m12} so that:
\be
m^2_{1,2}=\half m_{CS}^2 + e^2v^2 \pm \half m_{CS}^2 \sqrt{ 1+4
e^2v^2/m_{CS}^2}
\ee
In this case when $v=0$, the gauge boson contains one massive and
one
massless mode as expected for Maxwell-CS theory without symmetry
breaking.

The vertices are obtained from the interaction terms in the
Lagrangian.
They are shown in Fig. 1.  We use the following notation:  The dotted lines
are gauge bosons and all momenta are
ingoing.  Scalar lines are labelled $\chi$ or $\eta$ and the notation
$\chi/\eta$ indicates two separate diagrams.  The results are,
\bea
V_{1a} &=& -i\lambda;\,\,\,\,V_{1b}=-i\lambda/5 \none
V_{2a} &=& -i\lambda v;\,\,\,\,V_{2b} = V_{2c} = -i\lambda
v/5;\,\,\,\,
 \none
V_{3a} &=& -i\lambda v^2/2;\,\,\,\, V_{3b} = V_{3c} = -i\lambda
v^2/10;\,\,\,\,
 \none
V_{4a} &=& -i\lambda v^3/6;\,\,\,\,V_{4b} = -i\lambda v^3/30 \none
V_5 &=& e(P-Q)_\lambda \none
V_6 &=& i\gamma_0\epsilon^{\tau\alpha\lambda}K_\tau P_\alpha \none
V_7 &=& 2ie^2g_{\alpha\beta} \none
V_8 &=& -\delta \epsilon^{\alpha\lambda\beta}(P-Q)_{\lambda} \none
V_9 &=& -\delta v \epsilon^{\mu\alpha\nu}(K-Q)_\alpha \none
V_{10} &=& 2ie^2 v g_{\mu\nu}
\eea
\section {One Loop Effective Potential}
In this section we outline the calculation of the zero temperature
renormalized one loop effective potential.  When performing this calculation,
we want to consider the general
expression $\gamma=\gamma[v]$ in order to show that
renormalizability requires us to choose $\gamma=\gamma_0$, a constant.  To
obtain the one loop effective potential, we only need to know the propagators
and, as discussed earlier, the result for the photon propagator given
in the previous section is valid for arbitrary $\gamma=\gamma[v]$.
A straightforward calculation yields the Euclidean momentum space
integral for the
one loop effective potential:
\be
V^{(1)}(v)
= {\hbar\over 4\pi^2} \int dk k^2 \left(\ln(k^2+m_\chi ^2)
+\ln(k^2+m_\eta ^2)
+ \ln(k^2
+ m_1^2) + \ln(k^2 + m_2^2)\right)
\label{eq: effective potential integral}
\ee
Introducing a cut-off $\Lambda$, the basic zero temperature
integral that needs to be considered is (for $m^2>0$):
\be
\int^\Lambda_0 dk k^2 \ln(k^2+m^2) =
\frac{1}{3}\Lambda^3 \ln \Lambda^2 - \frac{2}{9}\Lambda^3
+\frac{2}{3}m^2\Lambda -{1\over3}\pi |m|^3
\label{eq: zero temp integral}
\ee
Using this result in \req{eq: effective potential integral} we find an
expression for
the complete one loop effective potential. Since \req{eq: m12} is valid
arbitrary $\gamma=\gamma[v]$, we obtain an expression for the effective
potential as a function of arbitrary $\gamma[v]$.
All masses are assumed
positive and from now on we drop the absolute value signs.  We obtain,
\bea
V(v) &=& {\lambda\over 6!} v^6+{\hbar\over 6\pi^2}(2e^2 v^2 + e^2
\gamma^2 v^4 +\lambda v^4/4! + \lambda v^4 / 5!) \Lambda\none
& & +{\Upsilon \over 2}v^2 + {\Omega \over 4!} v^4 + {\Xi \over 6!}
v^6 \none
& &- {\hbar \over 12\pi} (m_1^3 + m_2^3 + m_\chi ^3 + m_\eta ^3)
\label{eq: effective potential unrenormalized}
\eea
where we have dropped field independent terms and added the
counter-terms appropriate for scalar QED in 2+1 dimensions. Clearly
the theory will be one loop renormalizable only if
\be
\gamma^2 v^4 = a + bv^2 + c v^4 + d v^6
\ee
for some constants $\{a,b,c,d\}$. However, by inspection of the
last line in \req{eq: effective potential unrenormalized}, we see
that unless $d=0$, the effective potential will be unbounded below
for large $v$. Hence there is at most a three parameter family of
allowed Lagrangians. If we require $\gamma$ to be a sum
of rational functions of $|\phi|$, then
there are only two distinct possible solutions.
The first solution is obtained if $c,a =0$ which gives
$\gamma[v]=\sqrt{b}/v$.
This solution is used by Paul and
Khare \cite{pk}.  We note, however, that a CS term of this form
introduces an
explicit dependence on $|\phi|$ into a Lagrangian with other terms
that
depend only on $\phi^*\phi$ and therefore we will not consider this
solution.
The second solution is obtained if the right hand side of Eq.(23)
is a
perfect square which gives a solution of the form
$\gamma[v]=m_{CS}/ev^2 + \gamma_0$.  This solution is the
only one
that preserves the structure of the original Lagrangian.  As was discussed in
the Introduction, the first term in this expression gives the usual CS term.
Thus we have the remarkable result that $\gamma[|\phi|] = \gamma_0 =
constant$ is the unique non-minimal CS term that can be added to the usual CS
term while preserving one loop renormalizability of the effective potential.
\par
For
$\gamma_0=constant$, the mass and
quartic coupling can be renormalized to zero at $v=0$ by choosing:
\bea
\Upsilon &=& - {2\hbar e^2 \Lambda\over3\pi^2}\none
\Omega &=& -{\hbar\over 6\pi^2} (4!e^2\gamma_0^2 +
6\lambda/5)\Lambda
\label{eq: counterterms}
\eea
All the derivatives of the finite part of the effective potential to sixth
order
are zero when $v$ approaches zero from the right.  Therefore,
in contrast to the Coleman and Weinberg problem, all the
renormalizations can be done at $v$ equal to zero, and do not receive
contributions from the derivatives of the finite part of the effective
potential.
\par
As
noted by Coleman and Weinberg, self-consistency of the loop
expansion demands
that the one loop contributions proportional to $m_\chi ^3 \sim
m_\eta ^3 \sim
\lambda^{3/2}v^6$ be neglected compared to the
tree level $\lambda$. In this case, since the renormalized mass and
quartic
couplings are chosen to be zero, the remaining one loop effective
potential
is completely gauge and parametrization independent. The same
result is obtained in the unitary gauge in \cite{phys lett}.   One
way to understand
this is to note that all such problems occur in the scalar sector
of the
theory, which we are neglecting at the one loop level.
Equivalently, we note
that the effective potential is manifestly gauge invariant
because it is evaluated ``on shell'' \cite{nielsen}: when
$\lambda\approx 0,
\phi=constant$ solves the classical field equations.  Consequently,
the
Vilkovisky-DeWitt corrections used in \cite{burgess} to ensure
manifest gauge
fixing independence, for example, are identically zero.
\par
The gauge invariant renormalized one loop effective
potential is:
\bea
V&=&{\lambda v^6\over 6!} - {\hbar\over 12\pi}e^3 v^3 f(x)\none
&=&{\lambda \over 6! \gamma_0^6} (x^6- A x^3 f(x))
\label{eq: zero temp V}
\eea
where we have defined the dimensionless quantities $x=\gamma_0 v$
and
\be
A= {6!\hbar \over 12\pi}\left(\delta^3\over \lambda\right).
\label{eq: constant A}
\ee
In the above:
\be
f(x)= [1+\half x^2 + \half x \sqrt{4+x^2}]^{(3/2)} + [1+ \half x^2
- \half
x \sqrt{4+x^2}]^{(3/2)}
\ee
The potential in \req{eq: zero temp V} exhibits symmetry breaking
as long as the constant $A<1$. This is easily seen by noting first
that as $x\to0$, $V\sim - x^3$, while for large $x$, $V\sim (1-A)
x^6$. Thus, the requirement that the potential be bounded below
guarantees the presence of symmetry breaking, and puts the
following restriction on the ratio of dimensionless couplings in
the theory:
\be
{\delta^3\over \lambda} < {12\pi\over 6! \hbar}
\label{eq: bound delta}
\ee
Since consistency of the one loop approximation requires
$\lambda<1$, this also puts a corresponding bound on $\delta$ of
approximately $\delta < (1/20)^{1/3}$. In Fig. 2, the effective
potential
is plotted (up to an overall constant) for $A = 1/4$. The zero
temperature
potential is given by the curve labeled $g=\gamma_0 T/e=0$.  The
location of the symmetry breaking minimum $\xbar=\gamma_0\vbar$ is
easily found in terms
of the couplings. The implicit equation is:
\be
A= {6\xbar^3 \over (3 f(\xbar) + \xbar f'(\xbar))}
\ee
 A more useful form of the above is:
\be
{\lambda \over 5!}= {e^3 \over \vbar^3} {\hbar \over 12\pi}  (3
f(\xbar) +
\xbar f'(\xbar))
\label{eq: xbar}
\ee
In contrast to 4-d scalar QED, the scalar self-coupling cannot be
expressed solely in terms of the electromagnetic coupling constant,
since now the latter has dimensions of $L^{(-\half)}$.
\par
We can evaluate the mass ratios of the gauge bosons and scalar
particles.
The $\eta$ mass is $m^2_\eta(\vbar) = \lambda \vbar^4/5!$ and the
masses of
the gauge bosons are
given by $m^2_1(\vbar)$ and
$m^2_2(\vbar)$, where $m_1^2$ and $m_2^2$ are defined in \req{eq:
m12}.
The mass of the $\chi$ field is,
\bea
m_\chi^2 &=& \left. \frac{\partial^2 V}{\partial v^2} \right|_{\vbar} \none
         &=& \frac{\hbar e^3 \vbar}{12\pi} (9f(\xbar) - \xbar
f'(\xbar)
 -\xbar^2f''(\xbar))
\eea
Using $e/\vbar = \delta / \xbar$, we obtain,
\bea
{m_\chi^2 \over m_{1,2}^2} &=& \frac{\hbar\delta}{12\pi}
\left( \frac{9f(\xbar) - \xbar f'(\xbar)-\xbar^2f''(\xbar)}{\xbar(1+\half
\xbar^2 \pm
\half \xbar \sqrt{4+\xbar^2})}\right)
 \none
{m_\eta^2\over m_{1,2}^2} &=& \frac{\lambda}{5!\delta^2}
\left( \frac{\xbar^2}
{1+\frac{1}{2}\xbar^2 \pm \frac{1}{2}\xbar\sqrt{4+\xbar^2}}\right)
\label{eq: massratios}
\eea
Using $e/\vbar = \delta/\xbar$ in \req{eq: xbar}, we have $\xbar$
  as a function of $\lambda$ and $\delta$.  Thus, \req{eq:
massratios}
effectively expresses
the mass ratios as functions of these two dimensionless couplings,
as
required.  By expanding \req{eq: massratios} in
$\delta$, we obtain an expression that gives the effect of the
non-minimal CS term on the physical parameters.  We obtain,
\be
{m_\chi^2 \over m_{1,2}^2} = {3\over5!}\left( \frac{6!\hbar}{12\pi}\right)
^{2/3}\left( 1 \mp \delta \left(\frac{6!\hbar}{12\pi \lambda}\right) ^{1/3}
\right)
\label{eq: ratios}
\ee
The contribution from the non-minimal CS term has an appreciable effect on the
mass ratios when
\be
\frac{\delta^3}{\lambda}\sim \frac{12\pi}{6!\hbar}
\ee
We note that the contributions in \req{eq: ratios} become equal at
precisely the point at which the potential becomes unbounded when $x$
approaches infinity \req{eq: bound delta}.  Therefore, the non-minimal
CS term that we have introduced can never dominate the mass ratios.
{}From the lowest curve in Fig. 2 we obtain $\xbar=.87$,
 when $T=0$ and $A=1/4$.  This gives $m_{\chi}^2=0.7e^3\vbar$,
$m_1^2=2.33e^2\vbar^2$ and
$m_2^2=0.43e^2\vbar^2$.  The mass ratios are, $m^2_\chi/m_1^2\sim
0.34\delta$
and $m_\chi^2/m_2^2\sim 1.8\delta$.
\section{One Loop Renormalizability}
\par
In this section we discuss the one loop renormalizability of the
Lagrangian \req{eq: newlagrangian}, \req{eq:
gflagrangian}.
Power counting tells us that the superficial degree of divergence
of any
diagram is given by
\be
\Delta = 3-\half (E_B + E_S) -\half V_2 - V_3 -\frac{3}{2} V_4
-\half V_5
-\half V_6 -V_7 -\half V_9 -\frac{3}{2}V_{10}
\ee
where $E_B$ and $E_S$ are the number of external gauge boson and
scalar lines,
and $V_i$ indicates the number of vertices of the $i$th type, as
defined in
Fig. 1.  We note that this expression is only valid at one loop:
There are two
derivatives in the term in the Lagrangian that corresponds to $V_6$
but,
in any given one loop diagram, a vertex of the type $V_6$ can only
contribute
one power of internal momentum to the loop integral because of the
presence of
the epsilon tensor.  This argument is not valid for higher loop
diagrams
and thus our proof of renormalizability is only valid to one loop
order.
We use the standard BPHZ formulation of the renormalization
procedure \cite{IZ}.  We define the renormalized quantities in
terms of the bare quantities:
\bea
\chi_R &=& Z_\chi^{-\half} \chi \none
\eta_R &=& Z_\eta^{-\half} \eta \none
A^{\mu}_R &=& Z_A^{-\half} A^\mu \none
e_R &=& Z_e e \none
\eea
These definitions allow us to rewrite the original Lagrangian:
\be
{\cal L} = {\cal L}_R + {\cal L}_{CT} \none
\ee
${\cal L}_R$ is of the same form as the original Lagrangian with
the
bare quantities replaced by the renormalized ones.  The
counter-term
Lagrangian ${\cal L}_{CT}$ is obtained by calculating
the
contributions from divergent diagrams using the renormalized
Lagrangian and
constructing ${\cal L}_{CT}$, also as a function of renormalized
quantities, to cancel these divergences.  We use Pauli-Villiars
regularization to avoid the difficulties associated with the use of
dimensional regularization for the
calculation of
diagrams containing epsilon tensors.  From now on, all quantities
are the
renormalized ones, and the subscripts $R$ are omitted.
\par
To identify divergent diagrams we require $\Delta \geq 1$ since,
in 2+1
dimensions, all diagrams with $\Delta=0$ give zero by symmetric
integration.
In addition, there are many diagrams with $\Delta\geq 1$
which are, in fact, finite.  This effect occurs
when a dependence on external momenta reduces the actual degree of
divergence.
The identification of these diagrams is straightforward except for
the
diagrams in Fig. 3.
For these diagrams, in order to preserve gauge invariance, the
Pauli-Villiars
regulator must be applied to the closed scalar loop
and not to the individual propagators.  The resulting expression
is finite.
The technique is exactly analogous to that used for the closed
fermion loop contribution to the photon polarization tensor in
ordinary
spinor electrodynamics.
\par
The divergent diagrams are shown in Figs. 4-15. On each diagram,
the
vertices are labeled as in Fig. 1.  The dotted lines are gauge bosons, and
all momenta are ingoing.  The scalar lines are labeled
$\chi$ or $\eta$ and the notation $\chi / \eta$ indicates two separate
diagrams.
The results are given below, where the variable $M$ is the
Pauli-Villiars
mass.
\bea
{\rm Fig.\,\,\,\, 4a} &\rightarrow& ie^2vM/2\pi \none
{\rm Fig.\,\,\,\, 4b} &\rightarrow& i\lambda v^3 M / 40\pi \none
{\rm Fig.\,\,\,\, 4c} &\rightarrow& i\delta^2 v^3 M/2\pi \none
{\rm Fig.\,\,\,\, 5a} &\rightarrow& ie^2M/2\pi \none
{\rm Fig.\,\,\,\, 5b} &\rightarrow& 3i\lambda v^2 M / 40 \pi  \none
{\rm Fig.\,\,\,\, 5c} &\rightarrow& i\lambda v^2 M / 40\pi \none
{\rm Fig.\,\,\,\, 5d} &\rightarrow& i\delta^2 v^2 M / 2\pi \none
{\rm Fig.\,\,\,\, 5e} &\rightarrow& i\delta^2 v^2 M / \pi \none
{\rm Fig.\,\,\,\, 6a} &\rightarrow& 3i\lambda v M / 20\pi \none
{\rm Fig.\,\,\,\, 6b} &\rightarrow& 3i\delta^2 v M / \pi \none
{\rm Fig.\,\,\,\, 7a} &\rightarrow& i\lambda v M / 20\pi \none
{\rm Fig.\,\,\,\, 7b} &\rightarrow& i\delta^2 v M / \pi \none
{\rm Fig.\,\,\,\, 8a} &\rightarrow& 3i\lambda M / 20\pi \none
{\rm Fig.\,\,\,\, 8b} &\rightarrow& 3i\delta^2 M / \pi \none
{\rm Fig.\,\,\,\, 9a} &\rightarrow& i\lambda M / 20\pi \none
{\rm Fig.\,\,\,\, 9b} &\rightarrow& i\delta^2 M / \pi \none
{\rm Fig.\, 10a} &\rightarrow& 5M\delta
\epsilon_{\alpha\lambda\beta}
       P^\lambda / 12\pi \none
{\rm Fig.\, 10b} &\rightarrow& i\gamma_0^2 M
(g_{\alpha\beta}P^2-P_\alpha
       P_\beta) / 12\pi \none
{\rm Fig. \,10c} &\rightarrow& -i\delta^2 v^2 M g_{\alpha\beta} /
4 \pi
       \none
{\rm Fig.\,\,\,\, 11} &\rightarrow& -\delta \gamma_0  M (P_1^{\mu}
- P_2^\mu) /
       \pi \none
{\rm Fig.\,\,\,\, 12} &\rightarrow& -i\delta^2 M g_{\mu\nu} /2\pi
\none
{\rm Fig.\,\,\,\, 13} &\rightarrow& -i\gamma_0^2 M / 4\pi \none
{\rm Fig. \,\,\,\,14} &\rightarrow& -\delta\gamma_0 v M P_\mu / 4\pi
\none
{\rm Fig.\,\,\,\, 15} &\rightarrow& -i\delta^2  v M g_{\mu\nu}  /
2\pi
\label{eq: divergences}
\eea
To cancel these divergences, the counter-term Lagrangian must have
the
form
\bea
{\cal L}_{CT} &=& \half \alpha_1 (\partial_\mu \chi)^2
   + \half \alpha_2 (\partial_\mu \eta)^2
   + \half \beta_1 e^2 g^{\mu\nu}A_\mu A_\nu\chi^2
   + \half \beta_2 e^2 g^{\mu\nu}A_\mu A_\nu\eta^2
\none
&+& \nu eA^\mu(\chi \partial_\mu \eta - \eta \partial_\mu \chi)
+ \half \rho \chi  g^{\mu\nu}A_\mu A_\nu
   + \tilde{\gamma} e A^\mu\partial_\mu\eta
\none
&+& \half  z_1 A^\mu(\Box g_{\mu\nu} -
 \partial_\mu\partial_\nu)A^\nu
   + \half z_2 e^2v^2 g^{\mu\nu}A_\mu A_\nu
   + \half z_3 \delta v^2 \epsilon^{\mu\alpha\nu}
A_\mu\partial_\alpha A_\nu
\none
  & +& a\chi -\half \mu_1^2\chi^2 -\half \mu_2^2 \eta^2
   -\frac{\tau_1 }{4!}\chi^4-\frac{\tau_2}{4!} \eta^4
   -\frac{2\tau_3}{4!}\chi^2\eta^2 - \frac{\theta_1}{3!}\chi^3
   - \frac{\theta_2}{2}\chi\eta^2
\eea
The values of the coefficients in this expression are determined
from
the divergences as given in \req{eq: divergences}.  We have,
\bea
\alpha_1&=&\alpha_2=\beta_1=\beta_2=\nu=z_2 =
\frac{\gamma_0^2M}{4\pi}\none
\rho &=& \frac{\delta^2vM}{2\pi}\none
\tilde{\gamma} &=& \frac{\gamma_0 vM}{4\pi}\none
a&=&- \frac{M}{2\pi}[e^2v+\frac{\lambda }{20}v^3 + \delta^2
v^3]\none
\mu_1^2 &=& \frac{M}{2\pi}[e^2+\frac{3\lambda }{20}v^2 + 3\delta^2
v^2]\none
\mu_2^2 &=& \frac{M}{2\pi}[e^2+\frac{\lambda }{20}v^2 + \delta^2
v^2]\none
\tau_1 &=& \tau_2 = \tau_3 =
\frac{M}{2\pi}[\frac{3\lambda}{10}+6\delta^2]
 \none
\theta_1 &=& \frac{M}{2\pi}[\frac{3\lambda}{10}v + 6\delta^2
v]\none
\theta_2 &=& \frac{M}{2\pi}[\frac{\lambda}{10}v + 2\delta^2 v]
\label{eq: parameters}
\eea
To verify that the addition of the counter-term Lagrangian does not
introduce
any terms that were not present in the original Lagrangian, we
shift the real
part of the scalar field back to its original position:
$\chi\rightarrow
\chi - v$.  After making this shift, the counter-term Lagrangian
is,
\bea
{\cal L}_{CT} &=& \half \alpha_1(\partial_\mu \chi)^2
   + \half \alpha_2(\partial_\mu \eta)^2
+ \half \beta_1 e^2 g^{\mu\nu}A_\mu A_\nu\chi^2
   + \half \beta_2 e^2 g^{\mu\nu}A_\mu A_\nu\eta^2
\none
   &+& \nu eA^\mu(\chi \partial_\mu \eta - \eta \partial_\mu \chi)
   + \half  z_1 A^\mu(\Box g_{\mu\nu} -
\partial_\mu\partial_\nu)A^\nu
\none
   &+& \half z_3 \delta v^2 \epsilon^{\mu\alpha\nu}
A_\mu\partial_\alpha A_\nu
   + \half  g^{\mu\nu}A_\mu A_\nu[z_2 e^2v^2+\beta_1e^2v^2 - \rho
v]
\none
   &+& \chi g^{\mu\nu}A_\mu A_\nu[\half \rho - \beta_1e^2v]
   + A_\nu\partial^\mu\eta[ev\nu +e\tilde{\gamma}]
\none
   &+&\chi[a+v\mu_1^2 +\frac{\tau_1}{3!}v^3-\half\theta_1v^2]
   + \chi^2[-\half\mu_1^2 -\frac{\tau_1}{4}v^2 +\half\theta_1 v]
\none
   &+& \eta^2[-\half\mu_2^2 -\frac{\tau_3}{12}v^2 +\half\theta_2
v]
   + \chi^3 [\frac{\tau_1}{3!}v - \frac{1}{3!}\theta_1]
   + \chi\eta^2 [\frac{\tau_3}{3!}v-\half\theta_2]
\none
   &-& \frac{\tau_1}{4!}\chi^4- \frac{\tau_2}{4!}\eta^4
   - \frac{2\tau_3}{4!} \chi^2\eta^2
\eea
Substituting in the values for the coefficients as given by
\req{eq: parameters}
we obtain,
\bea
{\cal L}_{CT} &=& (Z_\phi^{-1} -1)\half |D_\mu \phi|^2
   - (Z_A^{-1} -1)\frac{1}{4} F_{\mu\nu} F^{\mu\nu}
\none
  &-& \frac{5M\delta}{24\pi}\epsilon_{\mu\alpha\nu}
A^{\mu}\partial^\alpha
 A^\nu
    -\frac{e^2M}{4\pi}\phi^*\phi -\frac{M}{2\pi
4!}(\frac{3\lambda}{10}+6\delta^2)
 (\phi^*\phi)^2
\label{eq: finalcounterterm}
\eea
where,
\be
Z_\phi^{-1} = 1+\frac{\gamma_0^2 M}{4\pi};\,\,\,\,\,\,\,\,\,\,
Z_A^{-1} = 1+\frac{\gamma_0^2 M}{12\pi};\,\,\,\,\,\,\,\,\,\,
Z_e = Z_A^{\,\,\,\, 1/2 }
\ee
We note that it is the non-minimal CS term that we have introduced that
generates the counter-term with the form of the standard CS term.
The last three terms in \req{eq: finalcounterterm} do not appear
in the
original Lagrangian because we explicitly set them to zero at the
tree level.
 They are analogous to the mass
renormalization term that appears in the Coleman-Weinberg
computation of
radiatively
induced spontaneous symmetry breaking for the $\phi^4$ theory with
zero bare
mass \cite{cw}.  These terms appear in our counter-term Lagrangian
because the
theory
does not possess a symmetry which guarantees the vanishing of the
bare scalar
mass, quartic scalar coupling and standard CS coupling
in the limit
where the renormalized values of these parameters are zero.
It is straightforward to see that the results would remain
unchanged if we had included non-zero bare scalar and CS mass terms
at the
tree level:  The
theory does not contain any diagrams with a degree of divergence
greater than
one and therefore, the divergence structure of the theory would not
change if
we had non-zero bare masses.
The only difference in the calculation would be a shift in the
values of the mass counter-terms which
are necessary to
give zero renormalized masses at $v=0$.  We obtain a renormalized
theory with
radiatively induced spontaneous symmetry breaking by setting the
physical
renormalized masses to zero at $v=0$ and using this equation as a
renormalization condition.
\section{Finite Temperature Effective Potential}
At finite temperature, $T$, the calculation of the effective
potential
in the Matsubara formalism proceeds by analytically continuing to
imaginary time
$t_E$ and imposing periodic boundary conditions at $t_E=0,
\beta=1/(kT)$. In \req{eq: effective potential integral}, the
integral over
$k_0$ is replaced by
a sum over discrete $k_0= 2\pi n T$ in the usual way
\cite{matsubara}. A
straightforward
calculation gives the temperature dependent part of the effective
potential:
\be
V_T(v) = {\hbar T\over 4\pi^2 } \int d^2 k \left[\ln( 1-
e^{-
\beta \omega_1}) + \ln (1-e^{-\beta \omega_2})\right]
\ee
where $\omega^2_{1,2}= \vec{k}^2 + m^2_{1,2}$.
We define the dimensionless variables $\theta=|\vec{k}|/T$,
$g=\gamma_0 T/e$ and
\be
y^2_{1,2}=\frac{m^2_{1,2}}{T^2}=
\frac{x^2}{g^2}\left(1+\frac{x^2}{2}\pm
\frac{x}{2}\sqrt{4+x^2}\right)
\ee
 Then,
\be
V_T(v) = \frac{\hbar T^3}{2\pi}\int d\theta\,\,
\theta \sum_{i=1,2} \ln (1-e^{-\sqrt{\theta^2+y_i^2}})
\label{eq: finite temp V}
\ee
Combining with \req{eq: zero temp V}, \req{eq: constant A} we have,
\be
V(v) = \frac{\lambda}{6!\gamma_0^6}(x^6-Ag^3\sum_{i=1,2}y_i^3 +
6Ag^3\int d\theta
\,\,\theta \sum_{i=1,2} \ln (1-e^{-\sqrt{\theta^2+y_i^2}})
\label{eq: total V}
\ee
Note that as in the zero temperature case we drop the contribution
from the scalar self coupling.
\par
The integral can be evaluated numerically, or we can find an
analytic
expression in the high
temperature limit.  Integrating by parts we can write,
\be
V_T(v) = -\frac{3\hbar T^3}{2\pi}\sum_{i=1,2} h_4[y_i]
\ee
where,
\be
h_4(y)= \frac{1}{\Gamma(4)}\int^\infty_0 {x^3\over
(x^2+y^2)^{\half}}
\left[\frac{1}{e^{\sqrt{x^2+y^2}}-1}\right] dx.
\ee
When $y < 1$ we can expand
\be
h_4(y)-h_4(0) = \frac{1}{12}\left[y^2 \ln y^2 - y^2\right] + {\cal
O}(y^4)
\ee
where we have dropped the terms $\sim y^3$ because they are
temperature
independent. This gives,
\be
V_T(v) = -\frac{\hbar T^3}{8\pi}\sum_{i=1,2}y_i^2(\ln y_i^2 -1 ).
\label{eq: expansion of finite temp V}
\ee
Combining \req{eq: finite temp V}, \req{eq: total V},
\req{eq: expansion of finite temp V} we obtain,
\be
V(v) = \frac{\lambda}{6!\gamma_0^6}\left(x^6-Ag^3\sum_{i=1,2}y_i^3
- \frac{3}{2}
Ag^3 \sum_{i=1,2}y_i^2 (\ln y_i^2-1)\right)
\ee
\par
When $y<1$
 this expression agrees with the result obtained from Eq.(45).
In Fig. 1 we show the numerical results for two different values
of $g$.
A first order phase transition is indicated.
\section{Conclusions}
We have introduced a non-minimal CS term
to massless scalar QED.
Renormalizability of the one loop effective potential puts severe
restrictions
on the form of the non-minimal coupling.  In particular, it was
shown that
only the term in Eq.(3) with with $\gamma=\gamma_0=constant$ can be added
to the
usual Chern-Simons Lagrangian without destroying the
renormalizability of
the effective potential at one loop.  It was shown that this theory
is fully
one loop renormalizable.
The purely massless theory is unstable due to radiative corrections
which
generate masses for all the particles through spontaneous symmetry
breaking.
At one loop, the symmetry is restored at high temperatures via what
appears to be a first order phase transition.  This phase transition is
qualitatively the same in the minimally coupled CS theory
however, unlike the standard CS term which gives only a shift in
the zero-point energy, the non-minimal CS term introduced in this paper does
affect the physics of the phase transition.
\par
Although the theory with $\gamma=constant$ appears to be quite
sensible at the
one loop level, more work is required to discover whether or not
higher loops
affect the results significantly.  As discussed in section 4, the power
counting analysis will
clearly not work at two loops in the same way that it does at one loop.
However, it is not impossible that the theory is renormalizable beyond one
loop.  Since the naive power counting argument indicates that the theory is
not renormalizable,
even at one loop, it is clear that power counting arguments can be misleading.
In addition, it is possible that the theory is a valid effective theory in
some restricted energy regime, and that in this regime the question of
renormalizability is not relevant.
Given the fact that the symmetry is
broken
radiatively in this (and other CS couplings) it would be
interesting to look
for vortex solutions \cite{jw} with the
given induced potential and CS coupling.  This work is in progress.

After the completion of this work, some additional references \cite{Stern},
\cite{Kogan} were brought to
our attention which motivate the physical interpretation of the coupling
$\gamma$ as a scalar magnetic moment.  This idea is discussed in detail in
\cite{ck}.

\par
\vspace*{20pt} {\Large\bf Acknowledgements} We are grateful to M.
Burgess, P.
Kelly, R. Kobes, M. Leblanc,  and D.J. Toms for useful discussions,
and to
Corissa Sweetland for help
with the diagrams.
This work was supported in part by the Natural Sciences and
Engineering Research Council of Canada.
\par

\vfill\eject
{\Large \bf Figure Captions}

\vspace*{1.0cm}
\noindent 1) Feynman rules for vertices \\
\noindent 2) The one-loop effective action for various values of
$g=\gamma_0 T/e$ \\
\noindent 3) A finite contribution to the scalar polarization tensor \\
\noindent 4) Some divergent contributions to the scalar one-point function \\
\noindent 5) Some divergent contributions to the scalar two-point function \\
\noindent 6) Some divergent contributions to the scalar three-point function \\
\noindent 7) Some divergent contributions to the scalar three-point function \\
\noindent 8) Some divergent contributions to the scalar four-point function \\
\noindent 9) Some divergent contributions to the scalar four-point function \\
\noindent 10) Some divergent contributions to the gauge boson two-point
function \\
\noindent 11) A divergent three-point function \\
\noindent 12) A divergent four-point function \\
\noindent 13) A divergent two-point function \\
\noindent 14) A divergent two-point function \\
\noindent 15) A divergent three-point function \\
\vfill\eject

\end{document}